\documentstyle{article}
\begin{document}
\newcount\nummer \nummer=0
\def\f#1{\global\advance\nummer by 1 \eqno{(\number\nummer)}
      \global\edef#1{(\number\nummer)}}
\def\nn{\nonumber \\}
\def\be{\begin{equation}}
\def\ee{\end{equation}}
\def\eel#1 {\label{#1}\end{equation}}
\def\ba{\begin{eqnarray}}
\def\ea{\end{eqnarray}}
\def\la{\label}
\def\re{(\ref}
\def\rz#1 {(\ref{#1}) }
\def\i{{\rm i}}
\let\a=\alpha \let\b=\beta \let\g=\gamma \let\d=\delta
\let\e=\varepsilon \let\ep=\epsilon \let\z=\zeta \let\h=\eta \let\th=\theta
\let\dh=\vartheta \let\k=\kappa \let\l=\lambda \let\m=\mu
\let\n=\nu \let\x=\xi \let\p=\pi \let\r=\rho \let\s=\sigma
\let\t=\tau \let\o=\omega \let\c=\chi \let\ps=\psi
\let\ph=\varphi \let\Ph=\phi \let\PH=\Phi \let\Ps=\Psi
\let\O=\Omega \let\S=\Sigma \let\P=\Pi \let\Th=\Theta
\let\L=\Lambda \let\G=\Gamma \let\D=\Delta

\def\w{\wedge}
\def\0{\over } \def\1{\vec } \def\2{{1\over2}} \def\4{{1\over4}}
\def\5{\bar } \def\6{\partial }
\def\7#1{{#1}\llap{/}}
\def\8#1{{\textstyle{#1}}} \def\9#1{{\bf {#1}}}

\def\({\left(} \def\){\right)} \def\<{\langle } \def\>{\rangle }
\def\[{\left[} \def\]{\right]} \def\lb{\left\{} \def\rb{\right\}}
\let\lra=\leftrightarrow \let\LRA=\Leftrightarrow
\let\Ra=\Rightarrow \let\ra=\rightarrow
\def\ul{\underline}

\let\ap=\approx \let\eq=\equiv 
        \let\ti=\tilde \let\bl=\biggl \let\br=\biggr
\let\bi=\choose \let\at=\atop \let\mat=\pmatrix
\def\CL{{\cal L}} \def\CD{{\cal D}} \def\rd{{\rm d}} \def\rD{{\rm D}}
\def\CH{{\cal H}} \def\CT{{\cal T}} \def\CM{{\cal M}} \def\CI{{\cal I}}
\newcommand{\dR}{\mbox{{\sl I \hspace{-0.8em} R}}}

\setcounter{page}{0}
\title{Quantization of Field Theories \\ Generalizing
Gravity-Yang-Mills Systems \\ on the Cylinder}

\author{P.\ Schaller\thanks{email: schaller@email.tuwien.ac.at, }
\ and T.\ Strobl\thanks{tstrobl@email.tuwien.ac.at}}
\date{Inst. f. Theor. Physik, Techn. Univ. Vienna, \\ Wiedner Hauptstr. 8-10,
A-1040 Wien, Austria\\[24pt] TUW-94-02, gr-qc/9406027}

\maketitle
\thispagestyle{empty}
\vfill
\begin{abstract}
Pure gravity and gauge theories in two dimensions are shown to be
special cases of a much more general class of field theories each of
which is characterized by a Poisson structure on a finite dimensional
target space. A general scheme for the quantization of these theories
is formulated. Explicit examples are studied in some detail. In
particular gravity and gauge theories with equivalent actions are
compared. Big gauge transformations as well as the condition of metric
nondegeneracy in gravity turn out to cause significant differences in
the structure of the corresponding reduced phase spaces and the quantum
spectra of Dirac observables. For $R^2$ gravity coupled to SU(2) Yang
Mills the question of quantum dynamics (`problem of time') is
addressed.

\end{abstract}

\section{Introduction}

Non abelian gauge theories as well as several models of gravity
on two dimensional space time manifolds have become an active
field of research in recent years
\cite{Raj},\cite{Het1},\cite{Het2},\cite{Gra},\cite{Kat},\cite{Ama}.
Both types of theories are closely related to each other. This
is best illustrated by the fact that the Lagrangian of a gravity
theory with
vanishing torsion and constant curvature (Jackiw Teitelboim
model \cite{JT}) can be rewritten as the one of a nonabelian
gauge theory with vanishing field strength ($BF$-theory)
\cite{Isl}.

Indeed, as shown in section 2 of the present article pure
gravity and gauge theories in two dimensions may be seen as
special cases of a  more general class of models each of which
is characterized by an antisymmetric tensorfield on a finite
dimensional target space $L$ (more precisely by a Poisson structure
on $L$).
A general scheme for the quantization
of these models in a Hamiltonian formulation (restricting the
topology of the space time manifold to the one of a cylinder) is
presented. The heart of this scheme is the reinterpretation of
the constraints as horizontality conditions on $U(1)$-bundles
over loop spaces. In the special case of non abelian gauge
theories the manifolds underlying the loop spaces are the
coadjoint orbits (the orbits generated by the gauge group on the
dual of its Lie algebra) equipped with the standard symplectic
structure \cite{KoS},\cite{Kir}.  In any case the result is a
finite dimensional quantum mechanical system generically
including discrete degrees of freedom of topological origin.

The considerations in section 2 are rather formal and abstract.
Explicit examples are given in section 3, including several
theories of quantum gravity. The reader may find it helpful to
have a look at these examples while reading section 2.

The constraints generate those symmetry transformations only
which are connected to the unity. The effect of large symmetry
transformations is studied in section 4 at the example of gauge
theories based on the $so(2,1)$ Lie algebra. We argue that the
quantum theory obtained by the implementation of the constraints
corresponds to an $\widetilde{SL}(2,{\bf R})$ gauge theory
($\tilde \cdot$ denoting the universal covering). For an
$SO(2,1)$ gauge theory the implementation of big gauge
transformations yields a one parameter family of unitarily
inequivalent quantum theories. The results obtained are checked
for both theories by investigating the topological structure of
the reduced phase spaces.  The latter are compared to the
reduced phase space (RPS) of the Jackiw Teitelboim model on a
cylinder characterized by the same action. Inequivalences found
are traced to the fact that the action of the constraints
generates diffeomorphisms only for space time manifolds with
nondegenerate metric and thus connect gravitationally
inequivalent solutions of the equations of motion.

In section 5 we investigate $R^2$-gravity coupled to an
$SU(2)$-Yang Mills theory. The Hilbert space and the operators
corresponding to a set of independent Dirac observables are
constructed  explicitly.  As in any quantum  theory of gravity
the Dirac observables are  space-time independent and the
Hamiltonian vanishes on physical quantum states.  Strategies to
resolve this apparent 'problem of (space-)time' \cite{Ish}  are
developed at the example of the reparametrization invariant
nonrelativistic free particle.  Realizing these strategies in
the gravity-Yang-Mills system, one
finds some partial confirmation of them through the fact that a gravity flat
limit reproduces the usual $SU(2)$ quantum dynamics.

The material covered in this work is based on two talks
delivered in Helsinki and St Petersburg
(cf.\ \cite{p6},\cite{p2},\cite{p5}).
To allow for a comprehensive treatment,
 more recent developments have been included as well
(cf.\ also \cite{p8p10}).

\section{The General Formalism}

The action $S$ of a non abelian gauge theory with a finite
dimensional semisimple gauge group is given by
\be
S=\int\langle F,*F \rangle , \quad F=dA+A\wedge A .
\eel actg
Here $\langle .,.\rangle$ denotes the Killing metric on the Lie
algebra of the gauge group, $A=A_\mu dx^\mu$ a Lie algebra
valued one form (gauge connection), and the Hodge dual $*$ is to
be taken with respect to a fixed metric on the space time
manifold.  If the latter is a cylinder $S^1\times {\bf R}$, one
may parametrize it by a coordinate $x^0\in{\bf R} $ and a $2\pi$
periodic coordinate $x^1$. For the Hamiltonian formulation one
may choose $x^0$ as the evolution parameter of the Hamiltonian
system. One then finds the zero component $A_0$ of the gauge
connection to play the role of a Lagrange multiplier giving rise
to the system of first class constraints (Gauss law constraints)
\be
\partial B(x^1) + ad^*_{A_1(x^1)}B(x^1)\approx 0  ,
\eel gaul
where $B$ is the momentum conjugate to the one component $A_1$
of the gauge connection and takes its values in the dual space
of the Lie algebra. The symbol $ad^*$ denotes the coadjoint
action of the Lie algebra on its dual space and $\partial$ the
derivative with respect to $x^1$.

All the physical systems considered in this paper have a
Hamiltonian structure generalizing the one of the non abelian
gauge theories.  They are obtained by modifying the constraints
(\ref{gaul}) due to
\be
G_i(x)=\partial B_i(x) + v_{ij}(B)A_1^j(x) \ap 0 \quad x\in S^1
\eel modc
where $B$ takes values in a linear space $L$, $A_1$ takes values
in the dual space $L^*$, and $v: L\to L \wedge L$ is a map from
$L$ into the space of antisymmetric tensors over $L$. [Indices
refer to an arbitrary basis in $L$ and the dual basis in $L^*$.
Summation over pairs of upper and lower indices is assumed.
Throughout this section we will abbreviate $x^1$ by $x$, as done
already in \re{modc}). The fundamental Poisson brackets are
given by $\{ A_1^i(x),B_j(y)\} = \delta^i_j\delta(x-y)$].  The
Gauss law of the nonabelian gauge theory is recovered from
\rz modc if $L^*$ is identified with the Lie algebra of the gauge group
and $v$ is chosen due to $v_{ij}(B)={f_{ij}}^k B_k$, where
${f_{ij}}^k$ denote the structure constants of the Lie algebra.

For general $v$, \rz modc will not define a system of first
class constraints.  Calculating the commutator of two
constraints, we find
\rz modc to be first class, iff
\be
{\6 v_{ij}\over\6 B_k}v_{kl} + cycl(i,j,l) =0 .
\eel conv
This is precisely the condition for $v$ to generate a Poisson structure on
$L$.\footnote{Cf., e.g., \cite{3Mad}; cf.\ also the article of A.\ Alekseev
and A.\ Malkin in the present Lecture Notes as well as \cite{p8p10}}.
The constraint algebra reads
\be \{G_i(x),G_j(y)\} = \delta (x-y) {\6 v_{ij}\over\6 B_k}G_k(x) .
\eel comc
The aim of the rest of this section is to investigate the
quantization of the system under the restriction \re{conv}).

To quantize the system in a momentum representation we consider
quantum wave functions as complex valued functionals on the
space $\Gamma_L$ of smooth parametrized loops in $L$:
\be
\Gamma_L=\{{\cal B}:S^1\to L, x\to B(x)\} .
\eel defg
Following the Dirac procedure \cite{Dirac}, we consider the
kernel of the quantum constraints
\be
\hat G_i(x)\Psi[{\cal B}]=\left(\partial B_i(x) + i\hbar
v_{ij}(B){\delta\over\delta B_j(x)}\right)\Psi[{\cal B}] =0 \eel
quac as the space $\cal H$ of physical states.

Let us consider two simple examples: For $v\equiv 0$ the support
of wave functions in $\cal H$ is restricted to constant loops.
Thus there is a natural identification of $\cal H$ with the
space of complex valued functions on $L$. If $v_{ij}(B)$ is a
constant invertible matrix, we may rewrite \re{quac}) according
to
\be
\left((v^{-1})^{ij}\partial B_j(x) -{\hbar \0 i}
{\delta\over\delta B_i(x)}\right)
\Psi[{\cal B}]=0
\eel rewc
and the physical wave functions have the form
\be
\Psi[{\cal B}]=c\, \exp\left({i \02\hbar}
\oint B_i(x)(v^{-1})^{ij}\partial B_j(x)dx\right)
, \quad c\in C .
\eel phyw
So in this case $\cal H$ can be identified with the complex
plane.

In general, $v$ is neither trivial nor nondegenerate.  For \rz
conv the vector fields
\be
V_i=v_{ij}(B){\partial \over \partial B_j}
\eel vecf
are in involution. Thus they generate an integral
surface\footnote{
possibly with singularities}
(symplectic leave) $I_{B_0}$ through any point $B_0 \in L$.  Denote by
$J =\{I_{B_0}, \, B_0 \in L
\}$ the space of these integral surfaces.
In $B_0 \in L$ the tangent vectors $V_i$ span a subspace
$S_{B_0}$ of the tangent space $T_{B_0}(L)$.  Given a cotangent
vector $w= w^idB_i \in T^*_{B_0}(L)$ in the kernel of $S_{B_0}$
(i.e.\ $v_{ij}(B_0)w^j=0$), we may use the antisymmetry of $v$ to
find ($B(x_0):=B_0$)
\be w^i \hat G_i(x_0)\Psi= w^i
\partial B_i(x_0)\Psi[{\cal B}]=0 . \la{ortr} \ee
Thus in any point $B_0$ the tangent vector $\partial B$ along a
loop ${\cal B}\in \Gamma_L$ is tangential to $S_{B_0}$, if
$\Psi[{\cal B}]\neq 0$.  In other words: The support of $\Psi$
is restricted to loops, which are entirely contained in some
integral surface $I \in J$.

Given a fixed element $I_0 \in J$, let us denote by
$\Gamma_{I_0}$ the space of loops on $I_0$. The vector fields
\be
{\cal V}_i(x)=v_{ij}(B(x)){\delta \over \delta B_j(x)}
\eel vecl
form an overcomplete basis in the tangent space over
$\Gamma_{I_0}$.  The ansatz $\Psi\vert_{\Gamma_{I_0}} =\exp\Phi$
for the restriction of $\Psi$ to $\Gamma_{I_0}$ allows to
rewrite the constraint equation according to
\be {\cal A}={\hbar \0 i}d\Phi
\eel equp
where ${\cal A}$ denotes the one form on $\Gamma_{I_0}$ given
implicitly by
\be
{\cal A}({\cal V}_i(x))=\partial B_i(x),
\eel desa
and $d$ denotes the exterior derivative on $\Gamma_{I_0}$.  The
above ansatz is general, if we exclude the trivial solution
$\Psi\vert_{\Gamma_{I_0}}\equiv 0$.  Locally eq. \re{equp})  is
integrable, iff $\cal A$ is closed. With the general identity
\be
d{\cal A}({\cal V}_i,{\cal V}_j)= {\cal V}_i\left({\cal A}({\cal
V}_j)\right)- {\cal V}_j\left({\cal A}({\cal V}_i)\right)
+ {\cal A}([{\cal V}_i,{\cal V}_j])
\eel ideo
and \rz conv we can indeed verify $d{\cal A}=0$, if the
constraints are first class.  Still, there could be global
obstructions to the integrability of
\re{equp}), if the first homotopy group of the underlying space,
$\Pi_1 (\Gamma_{I_0})$, is nontrivial. At this point one should
note, however, that ${\cal A}$ need not be exact, as $\Phi$ is
determined by $\Psi$ up to transitions $\Phi\to\Phi +i2\pi n$,
$n \in {\bf Z}$, only. Therefore $\Psi$ is well defined, iff
${\cal A}$ is integral, i.e.\ iff  ($h \equiv 2\pi\hbar$)
\be
\int_\gamma{\cal A}= n h, \quad n\in {\bf Z}
\eel intc
for any (noncontractible) closed loop $\gamma$ representing an
element of $\Pi_1(\Gamma_{I_0})$.  This condition yields a
restriction on the support of $\Psi$ to a (possibly discrete)
subset $\bar J$ of $J$.  For $I_0$ in this subset $\bar J$,
$\Psi$ is determined up to a multiplicative integration constant
on any connected component of $\Gamma_{I_0}$. (The space
$\Pi_0(\Gamma_{I_0})$ of connected components of $\Gamma_{I_0}$
is in one to one correspondence with the first homotopy group
$\Pi_1(I_0)$, \cite{loop}).
Let \be {\cal I} = \cup_{I\in {\bar
J}}\Pi_0(\Gamma_I) .
\eel tili
Then $\cal H$ is identified naturally with space of complex
valued functions on ${\cal I}$.

There is also a less abstract description of $\cal H$: Denote by
$\{Q_{(\a)}, \a=1,...,r\}$ a maximal set of
independent functions on $L$ invariant under the
action of the vector fields $V_i$.
We will denote those subspaces of
$L$, where the $Q_{(\a)}$ are constant, as their level surfaces
$M_Q$. If the connected
components of $M_Q$ are elements of $J$ (this is the generic
situation in many examples, c.f.\ next section),
the wave
functions can be written as
\be
\Psi[{\cal B}]=\widetilde{\Psi}(Q_{(r)},m_Q,n_Q)\,
\exp\left({i \0 \hbar} \int {\cal A}\right) ,
\eel wavf
where the discrete parameters $m_Q$, $n_Q$ characterize the
zeroth and first homotopy group of the level surfaces described
above.
\rz wavf yields the physical wavefunctions in terms of the variables
$B(x)$ and thus allows to describe the action of quantum
operators in $\cal H$.

An alternative formulation of the integrality condition \rz intc
is provided by the relation between one forms on a loop space
$\Gamma_M$ and two forms on the underlying space $M$: Any path
$\gamma$ in $\Gamma_M$ corresponds to a one parameter family of
loops in $M$ spanning a two dimensional surface $\sigma(\gamma
)$.  To any closed loop in $\Gamma_M$ the corresponding surface
in $M$ is closed.  Thus any two form $\o$ on $M$ generates a one
form $\alpha$ on $\Gamma_M$ via
\be
\int_\gamma\a = \int_{\s(\g)}\o
\eel conf
and $\alpha$ is closed and integral, iff $\o$ is closed and
integral. (The latter means that the integral of $\o$ over any
closed surface is an integer multiple of 2$\pi\hbar$).

Of course, not every one form on $\Gamma_M$ can be described in this
way.  In our case, however, the one form ${\cal A}$ on
$\Gamma_I$, $I\in J$ is generated by a two form $\O$ on $I$
characterized by its contraction with the vector fields
\re{vecf}):
\be
\O(V_i,V_j)=v_{ij} .
\eel cona
To prove this let us choose a path $\gamma \in \Gamma_I$
parametrized by a parameter $\tau \in [0,1]$. Any point in
$\gamma$ corresponds to a loop $B(x)$. Thus $\gamma$ induces a
map $S^1\times [0,1] \to L: (x,\tau)\to B(x,\tau)$. Denote by
$\dot B$ the tangent vector in the tangent space of $I$
corresponding to the derivative of this map with respect to
$\tau$. $\dot B$ as well as $\partial B$ can be written as
linear combination of the $V_i$:
\be
\dot B(x,\tau ) = \epsilon^i(x,\tau )V_i \, , \,
\quad \partial B(x,\tau ) = \mu^i(x,\tau )V_i \, \Rightarrow \,
\partial B_i=\mu^iv_{ij}.
\ee
The corresponding vectors in the tangent space of $\Gamma_I$
are now given by
\be
\dot {\cal B}
=\int_x\epsilon^i(x,\t){\cal V}_i(x) \, , \, \quad
\partial {\cal B}
=\int_x\mu^i(x,\t){\cal V}_i(x)  \,.
\la{linv} \ee
With \rz desa we have
\be
\begin{array}{rl}
\int_\g {\cal A}&=\int {\cal A}(\dot {\cal B})d\t =
\int\epsilon^i\partial B_idxd\t
=\\ &=\int\epsilon^iv_{ij}\mu^j dxd\t =\int\O(\partial B,\dot
B)dxd\t =
\int_{\s(\g)} \O \, .
\end{array}
\eel cali
Thus our assertion is proven. So \rz intc is equivalent to the
condition
\be
I \in  {\bar J}\quad\Leftrightarrow\quad \O \hbox{ is integral
on }I .
\eel nfin

$\O$ is an integral symplectic form and thus gives $I$ the
structure of an integrable space. Furthermore, $\O$ is invariant
under the flow of the vector fields $V_i$ (i.e.\ ${\cal L}_{V_i}
\O =0$, where $\cal L$ denotes the Lie derivative). So the
vector fields $V_i$ are locally Hamiltonian with respect to the
symplectic form $\O$ on $I$.

Let us illustrate the formalism by the example of non abelian
gauge theories (cf.\ also \cite{Ama}).  There L is the dual
space $g^*$ of the Lie algebra $g$ of the gauge group. Condition
\rz conv becomes the Jacoby identity. $J$ is the space of
coadjoint orbits, i.e.\ the space of orbits generated by the
action of the gauge group in $g^*$. The vector fields $V_i$ are
the vector fields on the coadjoint orbits associated with the
coadjoint action of the generators of the Lie algebra and
$\{Q_{(\a)}\}$ is the set of Casimirs on $g$.  $\O$ is the
standard symplectic form on the coadjoint orbits of a Lie group
as introduced in \cite{KoS}. The
coadjoint orbits are quantizable spaces, iff this symplectic
form is integral and their quantization yields the
unitary irreducible representations of $g$. This observation
establishes a connection of the momentum representation to the
configuration space representation of quantum mechanics for non
abelian gauge theories on a cylinder: In the configuration space
representation, where wave functions are functionals on the
space of gauge connections, the physical wave functions
(i.e.\ the kernel of the constraints) can be identified with
the functions on the space of unitary irreducible
representations of $g$ \cite{Raj}.

The generalization of our considerations to the case, where the
$B_i$ are local coordinates on a nonlinear space, is
straightforward. In this case $v \in T(L)
\wedge T(L)$ is an
skew symmetric two tensor over the tangent bundle and the
$v_{ij}$ are the components of $v$ with respect to the
coordinate basis in $T(L)$.  The formulation \rz quac of the
constraints can be made coordinate independent:
\be
 i_{df}\left[\partial B(x) + i \hbar v(B(x)) \right]\Psi[{\cal B}] = 0
\quad\forall\,f:L\to R .
\eel cooi

For $I\in {\bar J}$ and $B_0\in I$ denote by $v\vert_S(B_0)$ the
restriction of $v$ to $S(B)= T_{B}(I)$. We then have
$\O=(v\vert_S)^{-1}$. To prove this let us choose coordinates
$\{c_1,...,c_s,Q_{(1)},...,Q_{(r)}; s+r = dim L\}$ on $L$. In
these coordinates we have
\be v=\sum_{\a=1}^s v_{\a\b}{\delta\over\delta c_\a}
{\delta\over\delta c_\b}
\eel vres
as the $Q_{(\a)}$ are constant on $I$. Now \rz cona immediately
implies
\be
\O=(v^{-1})^{\a\b}dc_\a dc_\b .
\eel invv

\section{Explicit Examples}

The constraints \re{modc}) are induced by the action
\be\int [B_idA^i + {1 \0 2} v_{ij}(B) \, A^i\wedge A^j]   \eel udef
where $A$ is an $L^\ast$ valued one-form
and the zero-form $B$ is a map into the
corresponding dual space $L$. The action \re{udef}) is already
of  first order, i.e.\ Hamiltonian: $A_1$ and $B$ are seen to be
canonical conjugates  and $A_0$ enforces the constraints
\re{modc}).

The simplest nontrivial examples for  the considerations of the
previous section can be formulated in a three dimensional target
space $L$. For the rest of this section we will thus stick to
such a space. In three dimensions any antisymmetric two-tensor
$v$ can be rewritten according to
\be  v_{ij}(B) = \e_{ijk} u^k(B) \la{vij} \ee
for some $u^i$, $\e_{ijk}$ being the standard antisymmetric
$\e$-tensor with $\e_{123}=1$.
Indices may be raised and lowered by means of
the metric $\k_{ij}=diag(\pm 1,1,1)$.
The ${\epsilon_{ij}^k}$ can
be thought of as being the structure constants of $L^\ast :=so(3)$ or
$L^\ast := so(2,1)$
in an appropriate basis $\{T_i\}$.
If $u$ is the identity map, \re{udef}) takes the form $\int
B_iF^i$, which is the weak coupling limit of \re{actg}).

There are several further possibilities to  satisfy the generalized
Jacobi identity \re{conv}). One is provided by $u^k = u^k(B_k)$.
Another choice, of more interest  for the following, is given by
\be  u_a=B_a, \, u_3=u_3((B)^2,B_3), \quad (B)^2 \equiv B_aB^a,
\qquad
a \in \{1,2\} . \la{u}\ee Rewriting $A$ in the latter case as
\be A=e^aT_a+ \o T_3 \, ,\la{A}\ee
the action \rz udef takes the form
\be S_G= \int [B_a(de^a - \e^a{}_b \, \o \w e^b) + B_3d\o
+u_3(e^1\wedge e^2)] \, ,
\la{ga} \ee  with $\e_{ab}=\e_{ab3}$.
Much of our interest  in \re{ga}) stems from the fact that this
action can be reinterpreted as the action of a gravitational
theory: Viewing $e^a$ as a zweibein and $\o$ as a spin
connection,  the term following $B_a$ is identified with the
torsion two-form $De^a$, whereas $d\o$ becomes the  curvature
two-form.  $B_a$ and $B_3$ are vector  and scalar valued
functions, respectively, living on the two-dimensional manifold
characterized by the metric $g=e^ae^b \k_{ab}$. The latter is of
Euclidean or Minkowski type, corresponding to the respective
signature of the frame metric $\k_{ab}=diag(\pm 1,1)$.
Eliminating the $B$ fields by means of their equations of motion
for some special choices of $u_3$, we can establish the
equivalence of $S_G$ with purely geometrical actions of two
dimensional gravity.  E.g., $u_3=(1/4\g)(B_3)^2 -\l + {\a \over
2} (B)^2$ with $\a  \neq 0$ is easily seen to lead to 2D gravity
with torsion \cite{Kat}
\be  S^{KV}_G =- \int [\pm \g d\o \w \ast d\o \pm {1\0 2\a} De^a \w \ast De_a
+ \l \e], \label{KV} \ee
where $\e
\equiv e^1\wedge e^2$ is the metric induced volume form or
$\e$-tensor and '$\ast$' denotes the Hodge dual operation
($\ast \e =\pm 1$).
\re{KV}) is
the most general Lagrangian yielding
second order differential equations  for $e^a$ and $\o$ in two
dimensions. With $\a =0$ the same choice for $u_3$ leads to the
similar action of torsionless $R^2$ gravity.

To find the set $J$ of integral surfaces $I$, on which physical
wave functionals $\Psi[B]$ have their support, let us for a
moment return to an arbitrary dimensional target space $L$ and
a Poisson structure $v$ of the form $v = f_{ij}{}^k u_k(B) (\6/\6B_i)
\wedge  (\6/\6B_j)$ where the $f$'s are the structure constants of
some  Lie algebra of rank $r$; \re{vij}) may be regarded as a
special case of this.
 If $C(u) = C^{ijk...}u_iu_ju_k ...$ denotes one of the $r$
independent Casimirs,
\be w={\6 C(u) \0 \6 u_i} dB_i  \la{cas} \ee
will be  annihilated by $v$. In the case of a non abelian gauge theory
$u_i=B_i$ and $w=dC(B)$ so that according to \re{ortr}) the
(physical) wave functions $\Psi$ have support only on loops
$\cal B$ with  constant values of the Casimirs, as has been
noted already at the end of the previous section.  In the
present case of \re{vij}) the only independent Casimir  is the
Killing metric $\k_{ij}$, so that \re{ortr}) with \re{cas}, \ref{u})
becomes
\be (B^a\6B_a+ u_3\6B_3)\, \Psi =0. \eel aaaa
For reasons of calculational simplicity, let us specify $u_3$ to
\be u_3 = U(B_3)+ {\a \over 2} (B)^2. \la{U} \ee Multiplying now
\rz aaaa by the integrating factor $2\exp(\a B_3)$, we obtain
\be \6  Q \, \Psi = 0, \qquad Q = (B)^2 \exp(\a B_3) +
2 \int^{B_3} \! U(y)  \exp(\a y) dy \,. \la{q} \ee Thus
generically the level surfaces $M_Q$ generated by $Q= const$,
and thus (generically)  also the integral surfaces $I \in J$,
will be two-dimensional for the considered class of examples
\re{ga}).  In the following we will specify the potential $u_3$
to study these surfaces in more detail.

Prototypes  are  provided by the $SO(3)$ and  $SO(2,1)$
$BF$-theories based on the $so(3)$ and $so(2,1)$ algebra,
respectively, resulting from $u_3=B_3$.\footnote{Actually it is
rather the $BF$-theories of the corresponding universal covering
groups which have been quantized so far; the further steps necessary
to quantize an $SO(2,1)$-$BF$-theory will be discussed in the
following section.} Fixed values of the Casimir $Q \equiv
B^iB_i$ yield the coadjoint orbits of the groups, i.e.\ in the
former case spheres for $Q>0$ and the origin for $Q=0$ and in
the latter case one sheet hyperboloids for $Q>0$,  two sheet
hyperboloids for $Q<0$, and the light cone in the target space
for $Q=0$. In the compact case, we see that all the level
surfaces, which coincide with the integral surfaces, are two
dimensional except for the zero dimensional origin. The spheres
are connected and simply connected, but have $\Pi_2={\bf Z}$.
According to our general considerations of section 2 we
therefore know that the spectrum of $Q$ becomes discrete.  This
(i.e.\ eq.\ \re{intc}) or \re{nfin})) was a necessary  and
sufficient condition for the integrability of the horizontality
condition \re{quac}).  (The determination of this spectrum shall
be taken up at the end of this section, after having further
analyzed the topological structure of the integral surfaces).

In the noncompact $sl(2,{\bf R})$ case $\Pi_2$ is always trivial
and the spectrum of $Q$ remains continous.
For $Q<0$, however, the level surfaces $M_Q$ consist of two
(simply connected) parts, thus corresponding to two different
integral surfaces $I$ of the  vector fields $V_i$ defined in
\re{vecf}).  For $Q>0$ $\Pi_0(M_Q)$ is trivial, but
$\Pi_1(M_Q)={\bf Z}$; thus the integral surfaces $I$ of $V_i$
coincide with the level surfaces $M_Q$ in this case, but loops
$\cal B$ with different winding number  around the target space
hyperboloid are not smoothly connected to each other in the
space of loops on $I$ ($\Pi_0(\G_I) =\Pi_1(I) ={\bf Z}$); thus
for any winding number of ${\cal B} \sim B(x^1)$ we can
prescribe an independent initial value for the solution of the
first order differential equation \re{quac}). This illustrates
the necessity for the two quantum numbers $m_Q$ and $n_Q$ within
\re{wavf}). Actually, they correspond also to invariant Dirac
observables, if we allow the latter to become discontinuous:
Clearly
\be m_Q:= \Th(-Q) \Th(B_1), \qquad n_Q:= \Th(Q) \oint \6 \phi dx^1 \la{Dir} \ee
where $\Th$ is the Heaviside step function  and $\phi$ is the
angle variable of polar coordinates in the $(B_2,B_3)$-plane,
are  also Dirac observalbes in this extended sense, independent
from the continuous invariant $Q$.

The $Q=0$ level surface plays a special role: Since $V$ vanishes
at the origin, the latter  is an integral surface by itself and
splits $Q=0$ into three parts. (Note  that \re{quac}) constrains
the wave functionals to have support only on loops not passing
through a target space point where $V_i$ vanishes; thus this
splitting transfers consistently  to the spaces of loops on the
integral surfaces).  This implies also that $\Pi_1$ becomes
nontrivial for the future and the past target space light cones.
Allowing also for invariant distributions, we can uniquely
describe the integral surfaces of ${\cal V}_i \equiv V_i[B(x)]$,
i.e.\ the space $\cal I$ of eq.\ \re{tili}) (with $\bar J =J$
here), by means of Dirac observables:\footnote{The integral
surfaces of $V_i$ are characterized by the same quantities
except for $n_Q$.} Defining $\Th(0):=1$ we have to add merely
$\d[B_i]$ to $Q$, $m_Q$, and $n_Q$  so as to get a complete set
of independent commuting Dirac observables for the $sl(2,{\bf
R})$-$BF$-theory. The space $J$ of integral surfaces $I$ as well
as the space of loops on it, $\cup_{I\in J} \G_I$, are, however,
not Hausdorff at $Q=0$. As a consequence there might arise some
ambiguity in glueing together the orbit spaces with $Q\neq0$,
an issue which is certainly closely related to the determination
of an inner product.

Summing up the $sl_2$ case, we find the physical wave
functionals to effectively become functions (possibly also
generalized ones) on the space of the above  Dirac observables,
the corresponding spectra remain  classical, and the phase
factor becomes essentially superfluous, one  can get rid of it
by changing the basis in the U(1) quantum bundle.

Concerning the question of the inner product, let us remark only
that on large parts of the phase spaces of any of the models
\re{ga}) with \re{U}) and Minkowski signature, the  variable
conjugate to $Q$ can be written as ($e^\pm =: (e^2 \pm
e^1)/\sqrt{2}$)
\be P=-{1 \0 2} \oint \exp(-\a B_3) {e_1{}^-\0B_+}
dx^1 \approx -{1 \0 2}
\oint \exp(-\a B_3) {e_1{}^+\0B_-}      dx^1.  \la{P} \ee
Pulling through the phase factor of \re{wavf}), which in local
target space coordinates takes the form
\be \exp \left(- {i \0 \hbar} \oint\ln | B_+ |\6 B_3 dx^1 \right)
\sim \exp \left({i \0 \hbar} \oint \ln | B_- |\6 B_3 dx^1 \right) \ee
the
Dirac observable $P$ acts via $(\hbar/i) (d/dQ)$ on $\ti \Psi$.
Requiring that it  will become a Hermitean operator severely
restricts the measure of the inner product, but, in  the case
that $\ti \Psi$ depends also nontrivially on quantum numbers $m$
or $n$, this does not determine the inner product entirely. It
is not quite clear, if one should require  the 'Dirac
observables' $m_Q$ and $n_Q$, introduced above for the
$sl_2$-theory, to become hermitean as well. In this case the
corresponding eigenspaces would be orthogonal.

Next let us find the space of integral surfaces for
$R^2$-gravity, i.e.\ for \re{ga}) with potential $u_3=-(B_3)^2
-\l$, yielding $Q ((B)^2,B_3)= (B)^2 - 2(B_3)^3/3 -2\l B_3$.
For $\l >0$, $Q=const$ allows to determine $B_3$ uniquely as a
function of $(B)^2$.  Thus the resulting surfaces in the target
space are  diffeomorphic to a plane so that there is no
quantization of the classical spectrum of $Q$ and there are also
no additional quantum numbers within the wave functions
\re{wavf}). So for $\l >0$ the resulting Hilbert space is the
one of an ordinary particle on a line.

For $\l=0$ the situation is similar, only that the value $Q=0$
(critical value) plays a similarly exceptional role as in the
$BF$ case: one gets a conic singularity of the plane at
$(B)^2=B_3=0$ for Euclidean signature ($k_{ab}=\d_{ab}$), and
for Minkowski signature ($k_{ab} = diag(-1,1)$) additionally a
non Hausdorff structure (of $J$) at this point.

For $\l <0$ there are two  critical values of $Q$: $Q_{<(>)}
\equiv Q(0,\pm \sqrt{-\l}) = \pm 8(-\l)^{(3/2)}/3$, the  values
$\pm \sqrt{-\l}$ of  $B_3$ corresponding to the zeros of $u_3$
resp.\ $U$. For $Q \in (-\infty,Q_<) \cup (Q_>,\infty)$ the
resulting surfaces are again manifolds with trivial topology.
For $Q\in (Q_<,Q_>)$ and Euclidean signature  we get two
disconnected surfaces of the topology of a plane and a sphere,
respectively. Thus  the continuous spectrum  $Q \in {\bf R}$ has
a twofold degeneracy for some specific values of $Q \in
(Q_<,Q_>)$.  For $Q\in (Q_<,Q_>)$ and Minkowskian signature the
level surfaces $M_Q$ are connected  and of trivial second
homotopy; however, there are two fundamental noncontractible
loops, the winding numbers of which give rise to  a quantum
number $n_Q \in {\bf Z}$.

To analyse the situation for general potential $U$, it is
helpful to use a $(B)^2$ over $B_3$ diagram. Any fixed value of
$Q$ induces a  curve $C_Q$ in this diagram.  The intersections
of $C_Q$ with the $B_3$ axis are most crucial for
the topology of $M_Q$. Let us first consider the  Euclidean case,
where only non-negative values of $(B)^2$ are admissible: Any
part of $C_Q$ (in the positive of $(B)^2$) between two successive
intersections with the $B_3$ axis leads to a spherical $M_Q$,
any part of $C_Q$ with $(B)^2 \ge 0$ and exactly one point of
vanishing $(B)^2$ on it yields a 'plane', and a $C_Q$ with no
such points or intersections results in  a cylindrical $M_Q$ (or
an empty $M_Q$ for strictly negative $(B)^2$, as,  e.g., in the
$so(3)$-example for $Q<0$). Changes of the topology of $M_Q$
(along the choice of $Q$) can happen only at sliding
intersections of $C_Q$ with the $B_3$ axis; the latter are
possible only at $B_3=\b_c, \, U(\b_c)=0$, and thus only for the
'critical values' $Q_c=Q(0,\b_c)$ of $Q$.  The critical points
$(B_1,B_2,B_3)=(0,0,\b_c)$ (and only these) are then fixed points
of the vector fields $V_i$ and constitute an (zero dimensional)
integral surface by itself. For Minkowski signature the
transition from  $C_Q$ to $M_Q$  is a bit more cumbersome. The
result is, however, quite simple: If $C_Q$ contains no points
$(B)^2=0$, $M_Q$ consists of two disconnected 'planes'; if $C_Q$
contains $l$ points of (nonsliding) intersections with the $B_3$
axis, it has $l-1$  fundamental non-contractible loops.  At the
critical values $Q=Q_c$ (sliding intersections) we again have
fixed points $(0,0,\b)$, and the set $J$ of integral surfaces
becomes  non Hausdorff there.

For both signatures the fixed points correspond also to the
distributional solutions $\d[B_a]\d[B_3-\b_c]$ of the quantum
constraints and might be implemented via a point measure in the
inner product. (A somewhat special case arises when choosing
$u_3 \equiv 0$, describing 'flat gravity' on the cylinder, where
the set of $\b_c$ becomes uncountable and needs a separate
treatment).  Aside from these fixed point solutions the wave
functions have the form \re{wavf}). We further observe that in
our class of examples \re{ga}) the integral surfaces have a non
trivial second homotopy only for Euclidean signature and that
non trivial $\Pi_1$ implies trivial $\Pi_2$ and vice versa.

The discrete part of the spectrum of the Dirac observable $Q$ is
obtained most easily via the two-form $\O$ of section 2.
According to \re{invv}) it is the inverse of $v$ restricted to
the integral surfaces, which are (deformed) spheres in the case
under study. By construction $v(dQ, \cdot)=0$. Furthermore, due
to \re{vij}) and \re{u}) $v(dB_3, \cdot)$ is independent of the
potential $u_3$. Thus it will be convenient to calculate the
inverse of $v\mid_{M_Q}$ in coordinates $Q, B_3$ and e.g. $\ph =
\arctan (B_2/B_1)$; these cover the spheres up to the poles at
$B_a=0$. Since $v(dB_3,d\ph)=1$ we obtain
\be \O=dB_3 \wedge d\ph. \la{Two} \ee
Integrating this two-form over the considered 'sphere', the
integrability condition \re{intc}) becomes (cf.\ eqs.\
\re{nfin}, \re{conf}))
\be B_{3,max}(Q) - B_{3,min}(Q) = n \hbar , \qquad n \in {\bf N}  \la{N} \ee
where $B_{3,max}, B_{3,min}$ denote the values of $B_3$ at the
poles.  Given a curve $C_Q$ introduced above, it is then easy to
decide if this value of $Q$ allows for a spherical integral
surface or not.  For the case of $u_3=B_3$, \re{Two}) becomes
the rotation invariant Kostant-Souriau form $\O=rsin\dh d\dh
d\ph = (\e^{ijk} B_i dB_j dB_k /r^2)\mid_{r=\sqrt{Q}}$, where
$(r,\dh,\ph)$ denote spherical target space coordinates, and the
quantization condition \re{N}) can be expressed also explicitly
in terms of   $Q=r^2$, namely as $Q=n^2\hbar^2/4, \, n \in {\bf
N}$.  If we add to this $Q=0$, corresponding to the
distributional solution located at ${\cal B}=0$, this spectrum
coincides precisely with the one obtained in the connection
representation \cite{Het1}.

\section{Large Gauge Transformations and Metric NonDegeneracy}

The previous two sections have been devoted to the analysis of
the models under consideration in a Hamiltonian formulation,
where the symmetries of the system are expressed in terms of
first class constraints.  There are, however, some subtle points
connected with this approach:

- The constraints are the generators of infinitesimal symmetry
transformations. Large gauge transformations (i.e.\ symmetry
transformations not connected to the unity) cannot be generated
by infinitesimal transformations and thus they are not
determined by the constraints.

- In the  gravity theories presented in the previous sections
the zero components of the zweibein and the spin connection
played the role of Lagrange multiplier fields. We eliminated
them from the phase space as unphysical degrees of freedom. But
in a theory of gravity one usually requires the metric of the
space time mani\-fold to be nondegenerate (i.e.\ $\det g \neq 0$
everywhere).  Obviously it is difficult to realize this
condition after eliminating the zero components of the zweibein
from the phase space. Even if we allow for a degenerate metric,
the problem is not solved: The constraints \re{modc}) with
\re{u}) generate the symmetries of the gravity theory only for
$\det g \neq 0$ and turn out to connect gravitationally
inequivalent solutions separated in the phase space by regions
with a degenerate metric.

In the present section we will illustrate the importance of
these points by considering concrete  examples.  Our analysis
will include the explicit calculation of the reduced phase space
(i.e.\ the space of solutions of the equations of motion modulo
the symmetries of the model) for gauge and gravity theories
based on the $sl(2,R)$ Lie algebra. All of the theories
considered are characterized by the same Lagrangian
\be \int \langle B,F\rangle \la{bf} \ee
and thus a naive calculation of the constraints yields
equivalent Hamiltonian systems. Nevertheless we will find that
the reduced phase spaces differ as the symmetry contents of the
models differ.\footnote{This is similar to the inequivalence of the
symmetry generators $(d/dq)$ and $q(d/dq)$ on a line even when
disregarding $q=0$ \cite{p6}.}

In the first example let us regard the action (\ref{bf}) as the
one of a $PSL(2,R)$ gauge theory. $PSL(2,R)$ is the group
obtained from $SL(2,R)$ by the identification $1 \sim -1$ and is
isomorphic to $SO_e(2,1)$, the component connected with the
unity of $SO(2,1)$. Thus its Lie algebra is given by
\be  [T_i,T_j]=\e_{ij}{}^k T_k,  \la{alg} \ee
where the last index in the $\e$-tensor has been raised by means
of the Killing metric $\k_{ij}=diag(-1,1,1)$.
A possible matrix
representation of \re{alg}) is provided by the real matrices
$T_1 =i \s_2/2$, $T_2=-\s_1/2$, and $T_3=-\s_3/2$, where the
$\s_i$ are the Pauli matrices.  From this one finds
$\k_{ij}=2tr(T_iT_j)$ so that, e.g., the Dirac observable
$Q=B_iB^i$ introduced in eq.\ \re{q}) can be expressed
alternatively as $Q =2trB^2=-4detB$ ($B\equiv B_iT^i$).

The group $\cal G$ of symmetry transformations is the group of
smooth mappings from the cylinder into
$PSL(2,{\bf R})$:\footnote{There are no nontrivial principal
$G$-bundles on a cylindrical base manifold, iff the  chosen
structure (gauge) group $G$ is connected.}
\be
{\cal G}_{PSL(2,R)}=\{g:S^1\times {\bf R}\to PSL(2,{\bf R})\}
\eel gaut
The equations of motion,
\be
F=0, \qquad d B+[A,B]=0 , \la{eom} \ee yield the connection to
be flat and the Lagrange multiplier field $B$ to be covariantly
constant. Up to gauge transformations a flat connection $A$ on a
cylinder is determined by its monodromy $M_A={\cal P}\exp\oint A
\, \in PSL(2,{\bf R})$ generating parallel transport around the
cylinder ($\cal P$ denotes path ordering and the integration
runs over a closed curve $C$ winding around the cylinder once).
As the exponential map is surjective on $PSL(2,R)$, any
monodromy matrix can be generated by a connection of the form
$A=A_1dx^1$ where $A_1$ is constant:
\be
A=\left( \begin{array}{cc}  z &y+t \\ y-t&- z
\end{array}\right)dx^1, \quad t,y,z \in {\bf R}  \la{m} \ee
Constant gauge transformations act on $A$ via the adjoint action
leaving the determinant $t^2-y^2-z^2$ invariant and may be
interpreted as Lorentz transformations in the three dimensional
Minkowski space $(t,y,z)$.  Hyperbolic, elliptic and parabolic
elements, respectively, in the Lie algebra correspond to
spacelike, timelike, and lightlike vectors, respectively, in
this Minkowski space.  By Lorentz transformations in the
$(t,y,z)$ plane they can be brought into the form:
\be
\begin{array}{c}
A^{hyp}=\left( \begin{array}{cc}  0 & \a \\
\a& 0 \end{array}\right)dx^1, \quad
A^{ell}=\left( \begin{array}{cc}  0 & \dh\\ -\dh& 0
\end{array}\right)dx^1, \\ A^{par}=\left( \begin{array}{cc}  0 &
0 \\
\pm 1 & 0 \end{array}\right)dx^1
\end{array}\la{Arep} \ee
with $\a,\dh \in {\bf R} $ and the identification $\a\sim -\a$.
Exponentiation yields the monodromy matrices
\be \begin{array}{c}
M_{A^{hyp}}=\left( \begin{array}{cc}  \cosh 2\pi\a & \sinh
2\pi\a \\
\sinh 2\pi\a& \cosh 2\pi\a \end{array}\right), \quad
M_{A^{ell}}=\left( \begin{array}{cc} \cos 2\pi\dh  & \sin
2\pi\dh\\ -\sin 2\pi\dh& \cos 2\pi\dh \end{array}\right), \\
M_{A^{par}}=\left( \begin{array}{cc}  1 &  0 \\
\pm 2\pi & 1 \end{array}\right).
\end{array}\la{g} \ee
inducing the further identification $\dh\sim\dh +1/2$ in the
elliptic sector (remember $\oint dx^1=2\pi$ and $1 \sim -1$).
The integration of  the second eq.\ \re{eom}) gives
$B(x^0,x^1)=B(x^0,x^1+2\pi)=M_A B(x^0,x^1){M_A}^{-1}$ and thus
choosing a connection from (\ref{Arep}) $B(x)$ has to commute
with the corresponding monodromy matrix and consequently with
the connection itself.  Using \rz {eom} again one finds $B(x)$
to be constant. We obtain:
\be\begin{array}{c}
B^{hyp}=\left( \begin{array}{cc}  0 &c_1 \\ c_1& 0
\end{array}\right), \quad B^{ell}=\left( \begin{array}{cc}  0 &
c_2\\ -c_2& 0 \end{array}\right), \\ B^{par}=\left(
\begin{array}{cc}  0 &  0 \\ c_3 & 0 \end{array}\right),
\qquad c_i \in {\bf R} . \end{array}\la{mom}
\ee
In the case $A=0$ (corresponding to $\a =0$ or $\dh=0$,
respectively, in (\ref{Arep})) $B(x)$ is constant, too, but it
is not restricted by its commutator with the monodromy matrix.
It is, however, subject to constant gauge transformations, as
they leave $A=0$ invariant. Considerations similar to those
above show that also in this case gauge representatives of the
solutions are given by (\ref{mom}) with $c_3=\pm 1$ and the
identification $c_1\sim -c_1$.  (\ref{Arep}) and (\ref{mom})
give a complete parametrization of the reduced phase space of
the $PSL(2,{\bf R})$-gauge theory.

The group of gauge transformations ${\cal G}_{PSL(2,{\bf R})}$ as
defined above is not connected; rather it consists of  an
infinite number of components not smoothly connected to each
other: $\Pi _0({\cal G})=\Pi_1 (PSL(2,{\bf R})) = {\bf Z}$.  A
complete set of representatives for the components of ${\cal
G}_{PSL(2,{\bf R})}$ is given by
\be g_{(n)} = \left( \begin{array}{cc}
\cos (nx^1/2) & \sin (nx^1/2)\\
-\sin (nx^1/2) & \cos (nx^1/2) \end{array}\right) ,\qquad n\in
{\bf Z} .   \la{groupel} \ee Parametrizing the phase space as in
(\ref{Arep}) - (\ref{mom}) we also implemented these gauge
transformations. The action of the group elements $g_{(n)}$ on
the connections (\ref{Arep}) gives in the hyperbolic sector
\ba A^{hyp}_{(n)}&=& \left( \begin{array}{cc} \a\sin (nx^1)
 & \a\cos (nx^1) + n/2 \\
\a\cos (nx^1) - n/2 & -\a\sin (nx^1) \end{array}\right)dx^1
\nn
B^{hyp}_{(n)}&=&c_1\left( \begin{array}{cc} \sin (nx^1)  &  \cos
(nx^1)  \\ \cos (nx^1) & -\sin (nx^1) \end{array}\right)
.\la{trsol} \ea An analogous result is obtained in the parabolic
sector. In the elliptic sector the $g_{(n)}$ generate a
transformation $\dh \to \dh + n/2 $.  They are responsible for
the previous identification $\dh \sim \dh + 1/2$, which is removed
now.

With this knowledge it is straightforward to find the RPS for an
$SL(2,{\bf R})$ gauge theory: The action is the same as the one
of the $PSL(2,{\bf R})$ theory.  Gauge transformations of the type
$g_{(2l+1)}$, $l\in {\bf Z}$ are not allowed, as we do not have
the identification $1\sim -1$.  Consequently the hyperbolic
sector of the RPS is parametrized by ($A_{(0)}^{hyp}$,
$B_{(0)}^{hyp}$) and ($A_{(1)}^{hyp}$ $B_{(1)}^{hyp}$).  An
analogous result holds for the parabolic sector. In the elliptic
sector we have ($A^{ell}$, $B^{ell}$) but with the
identification $\dh \sim \dh+1$ rather than $\dh \sim \dh + 1/2
$. In contrast to the $PSL(2,{\bf R})$-case there are elements
of the RPS which cannot be represented by a constant connection.
This is a consequence of the non surjectivity of the exponential
map between Lie algebra and group in the case of $SL(2,{\bf
R})$.

 From the homotopical point of view $SL(2,{\bf R})$ is the double
covering of $PSL(2,{\bf R})$. Analogously, excluding all gauge
transformations not connected to the unity from $\cal G$ is
equivalent to choosing the universal covering
$\widetilde{SL}(2,{\bf R})$ as the gauge group of the theory.
(Note that $\Pi_0({\cal G}_{\widetilde{SL}(2,{\bf R})})=
\Pi_1(\widetilde{SL}(2,{\bf R})) =
\{1\}$). Thus the RPS of $\widetilde{SL}(2,{\bf R})$
is parametrized  by ($A_{(n)}^{hyp}$, $B_{(n)}^{hyp}$), $n\in
{\bf Z}$, $\a \in {\bf R}$, the analogous solutions in the
parabolic sector, and ($A^{ell}$, $B^{ell}$), $\dh \in {\bf R}$
(without any identification).

To see the significance of the  difference between the
$PSL(2,{\bf R})$ and the $\widetilde{SL}(2,{\bf R})$ gauge
theory, let us have a look on the quantization of the models: In
the first case the elliptic sector of the configuration space
(i.e.\ the space of gauge inequivalent connections) is compact
and thus we expect the possibility of unitarily inequivalent
quantum theories with a discrete spectrum for the momentum
operator (i.e.\ the Dirac observable  $Q  = -4\det B$). We may
compare this with the result obtained in the previous section:
There we used the Gauss law constraints to realize gauge
transformations in the quantum theory. As outlined above the
constraints generate those gauge transformations only which are
connected to the unity. So the quantum theory we obtained
corresponds to the  $\widetilde{SL}(2,{\bf R})$ gauge theory.
Indeed a continuous spectrum for $Q$ was found.  Furthermore, the
discrete parameter $n_Q$ within the wave functions is also
readily identified with the parameter $n$  of the hyperbolic
sector in the above parametrization of the RPS  of the
$\widetilde{SL}(2,{\bf R})$ theory.

To find the correct quantization of the $PSL(2,{\bf R})$ theory
we have to implement large gauge transformations. To this end
let us employ the exponential map in order to rewrite the Gauss
law. Starting from an initial loop ${\cal B}_0$ in some
coadjoint orbit any loop ${\cal B}$ may be written as ${\cal
B}=g{\cal B}_0g^{-1}$ for some $g=\exp X \, \in {\cal G}$,
$X:S^1
\to sl(2,{\bf R})$.
(As we mentioned above the exponential map is surjective on
$PSL(2,{\bf R})$).  If $g$ is connected to the unity, we also
have $g(t)=\exp tX \in {\cal G}$ for $t \in [0,1]$. The Gauss
law can then be rewritten as
\be
\oint \langle X,\partial(e^{tX}{\cal B}_0e^{-tX})\rangle dx^1 \,
\Psi[{\cal B}]
=i\hbar {\partial\over\partial t}
\Psi[e^{tX}{\cal B}_0e^{-tX}] .
\eel newg
With the identity
\be
\oint dx^1 \langle X, \partial(e^{tX}{\cal B}_0e^{-tX})\rangle =
- \oint dx^1 \langle e^{-tX}\partial Xe^{tX},{\cal B}_0\rangle =
- \oint dx^1 {\partial\over\partial t}\langle e^{-tX}\partial
e^{tX},{\cal B}_0\rangle
\eel lemm
integration over $t$ leads to
\be
\Psi[{\cal B}] =
\Psi[{\cal B}_0]\exp\left(\oint {i \0 \hbar}
\langle g^{-1}\partial g,{\cal B}_0\rangle dx^1\right),
\quad g\in {\cal G}_e .
\eel intg
An alternative derivation of this exponantiated form of the
Gauss law constraint is provided by a Fourier transformation of
the gauge invariance property of the physical wave functionals
in the connection representation \cite{JacGol}.  By construction
the wave functions calculated in the previous sections are the
general solutions of  eq.\ \re{intg}) for gauge transformations
connected to the unity. To quantize the $PSL(2,{\bf R})$ theory
we may rewrite $\cal G$ as the semidirect product of ${\cal
G}_e$ (the component connected to the unity) and the zero'th
homotopy group $\Pi_0({\cal G}) = {\bf Z}$:
\be
{\cal G}={\cal G}_e\times_s{\bf Z}
\eel semp
The most general incorporation of the second factor ${\bf Z}$
into the quantum theory will be to require the wave functionals
to transform according to a unitary representation $D_\th$ of
${\bf Z}$ characterized by an angle $\th$:
\be D_\th (n)=\exp  \left( {i2\pi n\th \0 \hbar}\right) . \la{repz} \ee
Taking together (\ref{intg}) and (\ref{repz}) we are thus lead
to
\be
\Psi[{\cal B}] \equiv \Psi[g{\cal B}_0g^{-1}]=
\Psi[{\cal B}_0]\exp\left[{i \0 \hbar} \left(\oint \langle g^{-1}\partial g
, {\cal B}_0 \rangle dx^1 + 2\pi n\th\right)\right]
\la{gaub} \ee
for $g$ in the $n$-th component of $\cal G$.

It is easy to verify that (\ref{gaub}) is compatible with group
multiplication, i.e.\ if it holds for two gauge transformations
$g_1$, $g_2$, it will also hold for the product $g_1g_2$.  The
group ${\bf Z}$ is generated by one element which may be
represented by $g_{(1)}$ as defined in (\ref{groupel}).  For
this reason the wave functions obtained in the sections 2 and 3
will solve (\ref{gaub}) for all $g\in \cal G$, if this identity
holds for $g_{(1)}$.  Furthermore (\ref{intg}) will hold for all
loops ${\cal B}$ which may be written as ${\cal B}=g^{-1}{\cal
B}_0g$ for some $g\in {\cal G}_e$, if it holds for the loop
${\cal B}_0$.  It is now obvious that in the hyperbolic sector
($Q>0$) the large gauge transformations simply relate the
$\tilde \Psi(Q,n_Q)$, ($n_Q \equiv n$), to each other for fixed
value of $Q$ and different values of $n$. In the elliptic sector
($Q<0$) we may apply (\ref{gaub}) to the constant loop ${\cal
B}_0=B^{ell}$ ($B^{ell}$ as defined in (\ref{mom})).  Noting
that $g_{(1)}$ commutes with $B^{ell}$, we find
\be {1 \0 \hbar}\left({1 \0 2\pi}\oint \langle g_{(1)}^{-1}\partial
g_{(1)},B^{ell}\rangle dx^1 + \th\right) = {\th -2c_2 \0 \hbar}
= {\th -sgn(B_1)\sqrt{-Q} \0 \hbar} \in {\bf Z}
\eel conq
in which the signum function can be expressed also in terms of
the quantum number $m_Q$ of eq.\ \re{Dir}) via $sgn(c_2)  \equiv
sgn(B_1) = 2m_Q-1$.  Thus in the elliptic sector the support of
physical wavefunctions $\tilde \Psi(Q,m_Q)$ is restricted to
$\sqrt{-Q}=(1-2m_Q)(l\hbar+\th)$, $l\in {\bf Z}$. The quantum
theories we obtain for different choices of $\th$ are obviously
unitarily inequivalent, as they generate  different spectra of
$Q$. This is precisely the result we expected.

At this point we want to mention that these results also
hold for the $PSL(2,{\bf R})$ Yang-Mills theory. In this case
$Q$ plays the role of the Hamiltonian.

Via the identification (\ref{A}) the action \re{bf}) together
with (\ref{alg}) may also be regarded as the one of a gravity
theory (Jackiw Teitelboim model). In this case the symmetry
content of the model is given by diffeomorphisms and local
Lorentz transformations (gravitational symmetries). This group
consists of a finite number of components not smoothly connected
to each other.  They differ by $x^0$- and $x^1$-reflection on
the space time manifold and by parity transformation and time
reversal in the Lorentz bundle. So up to these transformations
the symmetry content of the gravity theory seems to coincide
with the one of the $\widetilde{SL}(2,{\bf R})$ gauge theory.
So let us see, how the infinitesimal generators of the
gravitational symmetries are identified with the Gauss law
constraints (\ref{gaul}) generating the $sl(2,{\bf R})$ algebra.
To this end we may calculate the Hamiltonian density $\aleph$
($H=\oint \aleph dx^1$) of the theory as the generator of
diffeomorphisms in $x^0$-direction. We find $\aleph=- A_0^iG_i$,
with $G_i=\partial B_i + \e_{ij}{}^k A^j B_k$. Analogously the
generator of diffeomorphisms in $x^1$-direction is obviously
given by $A_1^iG_i$. Noting that $G_3$ precisely generates local
Lorentz transformations one concludes that identification of the
$sl(2,{\bf R})$ generators and the infinitesimal generators of
the gravitational symmetries crucially depends on the condition
$\det e \neq 0$.  Let us investigate the consequences of this
observation for the RPS of the gravity theory: With the
identifications \re{A}) the solutions we used above to
parametrize the RPS of the $\widetilde{SL}(2,{\bf R})$ gauge
theory correspond to space time manifolds with $\det g =0$. To
any of these solutions, however, it is possible to find a gauge
transformation yielding a solution corresponding to a
nondegenerate space time metric. More precisely, this can be
done in an infinite number of gravitationally inequivalent ways.
E.g., in the elliptic sector, we might apply one of the
following gauge transformations to $A^{ell}$ \ba & g_{[k]} =
\left(
\begin{array}{cc} \cos \c_k & \sin \c_k \\ -\sin \c_k & \cos
\c_k \end{array}\right) \left( \begin{array}{cc} 1 & b_k \\ 0 &
1 \end{array}\right) & \nn & \!\! \c_k = [\exp
(x^0)\,+2\vert\dh\vert ]\sin(kx^1), \, b_k =
[\exp(x^0)\,+2\vert\dh\vert ]\cos(kx^1) \quad k\in {\bf N} .  &
\la{smallgrel}
\ea
We obtain
\be A^{ell}_{[k]} = \left( \begin{array}{cc}
    b_k(\dh dx^1+d\c_k ) & (1+{b_k}^2) (\dh dx^1+d\c_k )+db_k \\
-(\dh dx^1+d\c_k ) & -b_k(\dh dx^1+d\c_k ) \end{array}\right) .
\la{novang} \ee The gauge transformations (\ref{smallgrel}) are
smoothly connected to the unity for arbitrary value of $k$ as
the $\c_k$ are periodic functions in $x^1$. Nevertheless the
solutions $A^{ell}_{[k]}$ are gravitationally inequivalent for
different values of $k$. To prove this let us again choose a
loop $C$ running around the cylinder once.  Under the
restriction $\det g = -(\det e)^2 \neq 0$ the components of the
zweibein $(e_0{}^+, e_1{}^+)$  induce a map $C\sim S^1\to {\bf
R}^2\backslash {\{0\}}$ characterized by a winding number (not
depending on the choice of $C$).  Solutions with different
winding numbers cannot be transformed into each other by
gravitational symmetries, since they are separated by solutions
with $\det e =0$. (Also the discrete gravitational symmetry
transformations mentioned above do not change the winding
number).  For different values of $k$ the solutions
(\ref{novang}) have different winding numbers, which proves our
assertion.

This result generalizes to the other sectors of the theory:
Solutions which are gauge equivalent in the
$\widetilde{SL}(2,{\bf R})$ gauge theory are not equivalent in
the gravity theory, if they have different winding number.

The winding number defined above is related to the kink number
as defined in \cite{kink} by means of 'turn arounds' of the
light cone along non contractible loops. More precisely, winding
number $k$ corresponds to kink number $2k$. (Odd kink numbers
\cite{kink} characterize solutions which are not time
orientable. Such solutions are not considered here).

The physical relevance of solutions with nontrivial winding
number is not quite clear. They necessarily contain closed
lightlike curves.  There are, however, also solutions with
trivial winding number containing closed lightlike curves. As
outlined, in a conventional Hamiltonian treatment of the action
\re{bf}) the constraints will generate infinitesimal gauge
transformations rather than gravitational symmetry
transformations. Thus on the Hamiltonian level the kink number
will not appear in the parametrization of the reduced phase
space, while, however, not all solutions with closed timelike
curves can be excluded in this way. A similar situation occurs
also when treating other models of two dimensional gravity
contained in \re{ga}). It would be interesting to see, if the
equivalence up to $\det g =0$ of the Hamiltonian and Lagrangian
formulation of four dimensional gravity leads to similarly
inequivalent factoring spaces.

\section{A Model for Quantum Gravity}

In the previous sections we found that a large class of
Hamiltonian systems, including gravitational ones, can be
reduced to quantum systems of finitely many topological
degrees of freedom. The question arises: Can such models serve
as toy models for a quantum theory of four  dimensional gravity?
Indeed even in the absence of local degrees
of freedom an illustrative treatment of some conceptual
questions of quantum gravity is possible.  Most prominent among
these  is the so called 'problem of time' \cite{Ish}, which we
shall take up in this section for the example of $R^2$-gravity
with Minkowski signature coupled to $SU(2)$ Yang Mills.

The Lagrangian of this system is
\be S=  \int_{S^1 \times {\bf R}} [{1 \0 8\b^2} R_{ab} \wedge \ast R^{ab}
+{1 \0 4\g^2} tr(F \wedge \ast F)]  \la{S}    \ee where   the
Hodge dual operation is, in contrast to \re{actg}), performed
with the dynamical metric used to define also the torsionless
curvature two-form $R_{ab}$, and the trace is taken, e.g., in
the fundamental representation of $su(2)$
(the generators $T_i$,
fulfilling \re{alg}) with $\k_{ij}=\d_{ij}$, are then
represented by $T_i=-i\s_i/2$, which yields $\k_{ij}
=-2tr{T_iT_j}$ now).
Rewriting $\re{S})$ by means of Cartan
variables $(\o^a{}_b :\equiv -\e^a{}_b \o, e^a)$ in a
Hamiltonian first order form, it becomes \be S_H=\int_{S^1
\times {\bf R}} B_aDe^a+B_3d\o + tr(EF) + [-\b^2 (B_3)^2+\g^2
tr(E^2)]\e
\la{SH} \ee
where we have chosen $E =E^iT_i$ to denote the 'electric fields'
conjugate to the $SU(2)$-connection one-components $A_1$, and
the $B$'s are the conjugates to the spin connection $\omega_1$ and the
zweibein one-components $e_1{}^a \equiv
(e_1{}^-,e_1{}^+)$.  (Our conventions are $e^\pm
=(e^2 \pm e^1)/\sqrt{2}$, yielding a light cone frame metric
$\k_{+-}=1$, whereas $\e=e^1 \wedge e^2= e^-\wedge e^+$ so that
$\e^{+-}=\e_{-+}=1$). Obviously $S_H$ is the sum of an
$SU(2)$-$EF$-theory \re{bf}) (up to a factor $-2$) and an action
$S_G$ \re{ga}) with $u_3= -\b^2 (B_3)^2 +\g^2 tr(E^2)$. In
explicit terms the constraints following (naturally) from $S_H$
are
\ba
G_a&=& \6 B_a + \e^b{}_a B_b \o_1 + \e_{ab}[-\b^2(B_3)^2+\g^2tr
E^2]   e_1{}^b  , \la{Ga}\\ G_3&=& \6 B_3 + \e_b{}^aB_a e_1{}^b,
\la{G3}
\ea
beside the unmodified $SU(2)$ Gauss law $G \approx 0$.  We will
not attempt to reformulate these constraints so as to possibly
cure the global deficiencies of them with respect to
diffeomorphisms  noted at the end of the previous section.
Instead we proceed with a straightforward quantization.

There are two independent Dirac observables as functions of the
momenta ($q_{(s)} \equiv \oint Q_{(s)} dx^1/2\pi$)
\ba q_{(1)}&=& {-1 \0 \pi} \oint tr(E^2) dx^1 \equiv {1 \0 2\pi}
\oint E_iE_i dx^1 \nn
q_{(2)} &=& {1 \0 2\pi}
\oint [(B)^2 -{2 \0 3} \b^2 (B_3)^3 +2\g^2 tr(E^2)B_3]dx^1 .  \nonumber \ea
The corresponding level surfaces have topology $S^2 \times {\bf
R}^2$ for $q_{(1)} \neq 0$ and ${\bf R}^2$ for
$q_{(1)}=0$.\footnote{Within the latter level surface  the
origin is an integral surface by itself.  We will in the
following disregard this small complication. -- As suggested
already through the chosen notation we assume $\b^2$ and $\g^2$
to be non negative.} This gives rise to the quantization
condition (cf.\ end of sec.\ 3): $q_{(1)} = n^2/4, \, n \in {\bf
N}_0$.  Expanding the physical  wave functionals in terms of
eigenfunctions of $q_{(1)}$, the corresponding coefficients are
\be  \exp \left({i \0 \hbar} \oint (E_3 \6 \varphi \pm \ln B_\mp \, \6
B_3 dx^1\right)  \ti \Psi_n(q_{(2)}),  \quad\,  n =
2\sqrt{q_{(1)}} \, \in {\bf N}_0, \,\, q_{(2)} \in {\bf R},
\la{psi} \ee where we have written the phase factor in some
local target space coordinates with $\tan \varphi \equiv (E_2/
E_1)$.  The inner product with respect to $q_{(2)}$ is
determined by the hermiticity requirement on \be p_{(2)} = - {1
\0 2} \oint {e_1{}^\pm \0 B_\mp}dx^1,   \la{p} \ee the Dirac
observable conjugate to $q_{(2)}$:  as noted already in section
3, $p_{(2)}$ acts as the usual derivative operator on $\ti
\Psi_n$, thus leading to the ordinary Lebesgue measure
$dq_{(2)}$.

We end up with the Hilbert space ${\cal H}$ of an effective
two-point particle system with  nontrivial phase space topology.
As a basic set of operators acting in $\cal H$ we could use
$q_{(2)},p_{(2)}$, $q_{(1)}$, and $tr [{\cal P} \,\exp (\oint
A_1dx^1)]$.
 From the latter one may construct a ladder operator $l:n\to n+1$.

All operators acting in $\cal H$ are thus found to be
expressible in terms of $q_{(2)},p_{(2)}$, and the number and
ladder operators. However, we do not have an operator such as
$g_{\m\n}(x^\m)$.  Following, furthermore, any  textbook on
elementary quantum mechanics, the next step in the quantization
procedure would be to introduce an evolution parameter 'time',
which we will call $\t$, and to require the wave functions to
evolve in this parameter according to the Schroedinger equation.
In the present case, however, the Hamiltonian following from
\re{SH}) is a combination of the constraints, \be H=-\oint
[e_0{}^aG_a + \o_0 G_3 +tr(A_0 \, G)],
\la{H} \ee so that the naive Schroedinger equation becomes
meaningless.

Both of these items, the nonexistence of space-time dependent
quantum operators as well as the apparent lack of dynamics, are
correlated and they are not just a feature of the topological
theory \re{S}). Also in four dimensional gravity the quantum
observables are some (not explicitly space-time dependent)
holonomy equivalence classes and the Hamiltonian vanishes when
acting on physical wave functions  \cite{Ash}. Diffeomorphisms
are part of the symmetries of any gravity theory; as a
consequence the Lie derivative into any 'spatial' direction can
be found to equal  the  Hamiltonian vector field of some linear
combination of the constraints (in our case ${\cal L}_1
=e_1{}^aG_a +\o_1 G_3 +tr A_1 \, G$), whereas, on shell,
$x^0$-diffeomorphisms will be generated by the Hamiltonian $H$.
Thus, although 4D gravity has local degrees of freedom, any of
its (uncountably many)  Dirac observables will be also
space-time independent.

To orientate ourselves as of how to introduce  quantum dynamics
within such a system, let us have recourse to the simple case of
a nonrelativistic particle (NRP). As is well known, any
Hamiltonian system can be reformulated in time reparametrization
invariant terms.  In the case of the NRP,
\be \int (p {dq \0 dt} - {p^2 \0 2})dt =\int (p \dot q  -
{p^2 \0 2} \dot t)  d\t, \la{rep} \ee the equivalent system has
canonical coordinates $(q,t;p,p_t)$ and the 'extended'
Hamiltonian  is proportional (via a Lagrange multiplier) to the
constraint $C=p^2/2 + p_t \approx 0$.  Quantizing this system,
e.g., in the coordinate representation, we observe that the
implementation  of the constraint $C\psi(q,t)=0$ is equivalent
to the Schroedinger equation of the original formulation, if one
reinterpretes the canonical variable $t$ as evolution parameter
$\t$. Therefore, given this formulation of the NRP or similarly
of any other system, the postulate of a Schroedinger equation
within the transition from the classical to the quantum system
becomes superfluous; rather it is already included within the
Dirac quantization procedure in terms of a constraint equation.

The identification $t =\t$ above can be looked upon also as a
gauge condition with gauge parameter $\t$.  This interpretation
is helpful for the quantization of the parametrization invariant
NRP in the momentum representation, in which case the space of
physical wave functions is isomorphic to the space of functions
of the Dirac observable $p$. The gauge condition $\bar C \equiv
t-\t =0$ provides a perfect cross section for the flow of $C$.
Thus it is possible to determine any phase space variable in
terms of the Dirac observables $p$, $Q=q-pt$, as well as the
gauge fixing parameter $\t$. Interpreting $\t$ as a dynamical
flow parameter 'time', the obtained evolution equations for $p$
and $q$, transferred to the quantum level as $q(\t)=i \hbar \,
d/dp +   \t p$, $p(\t)=p$,  become equivalent to the Heisenberg
evolution equations of the parametrized  NRP.

The operator $q(\t)$ above  corresponds to a measuring device
that determines the place of the particle at time $\t$.  A
measuring device that determines the time $t$ at which the
particle is at a given point $q=q_0$, on the other hand,
corresponds to the alternative gauge condition $\ti C \equiv
q-q_0 =0$. $\ti C$ provides a good cross section only for $p
\neq 0$. Ignoring this subtlety, e.g. by regarding only wave
functions with support at $p \neq 0$, the (hermitian) quantum operator for
such an experiment is $t(q_0)=-i \hbar \,[(1/p) d/dp -(1/2p^2)]
+ q_0/p$.  In this second experimental setting Heisenberg's
'fourth uncertainty relation' between time $t$ and energy $p^2/2
\sim -p_t$, usually motivated only heuristically, becomes a strict
mathematical equation. We learn that different experimental
settings are realized by means of different gauge conditions,
and, at least in principle, vice versa.

The wave functions of \re{SH}) are basically functions of the
Dirac observables, although  part of the latter  became
discretized in the quantum theory.  Transferring the ideas above
to the gravity system, we should find gauge conditions to the
constraints \re{Ga}, \ref{G3}).  (It will not be necessary to
gauge fix also $G$).  As such we will choose
\be \6 B_+=0, \quad B_3  + \t B_+ =0, \quad e_1{}^- =1.  \la{eich} \ee
It is somewhat cumbersome to convince oneself that this is
indeed a good gauge condition. However, for $q_{(1)} \neq 0$  it
provides even a globally well-defined cross
section.\footnote{One possibility to check the obtainability of
\re{eich}) is to carefully analyse the Faddeev matrix, taking
into account that due to $\oint \6  Q dx^1 \equiv 0$ and
\re{ortr}, \ref{q}) the gravity
constraints are not completely linearly independent.  This
(infinite dimensional) matrix turns out to be nondegenerate, iff
$B_+ \oint e_1{}^- dx^1 \neq 0$. For $q_{(1)}
\neq 0$ any gauge orbit in the loop space contains a
representative fulfilling this condition, which suffices to
prove the assertion since the space of gauge orbits is connected
in the case under study (no quantum number $n_Q$).} The gauge
conditions together with the constraints allow to express all
gravity phase space variables in terms of Dirac observables.  In
this way one obtains evolution equations such as
\be B_-(\t)=-{1 \0 2\pi} p_{(2)}q_{(2)} -{\g^2 \0 2}
q_{(1)} \t - {\b^2\pi^2
\0 3 (p_{(2)})^2}\t^3 ,\quad B_+(\t)={-\pi \0 p_{(2)}}. \la{B-} \ee
Antisymmetrizing this with respect to $q_{(2)}$ and $p_{(2)}$,
\re{B-}) can be taken as an operator in the Hilbert space $\CH$
defined above.\footnote{The elementary procedure above coincides
with the use of Dirac brackets for $\t$-dependent systems (in
which case one extends the symplectic form by $d\t \wedge
dp_\t$); this explains also $B_-$ and $B_+$ do not commute
anymore.} Similarly one finds  $g_{11}(x^0)=2e_1{}^+(x^0)=
-p_{(2)} B_-(x^0)/\pi$, $(x^0\equiv \t)$, which now, up to
operator ambiguities, becomes a well defined operator in our
small quantum gravity theory, too.

Requiring that the $\t$-dependence of \re{eich}) is generated by
the Hamiltonian $H$, the gauge conditions determine also the
zero components of the zweibein and the spin connection.
Actually, one zero mode of these Lagrange multiplier fields
remains arbitrary as a result of the linear dependence of the
constraints $G_i$ (cf.\ also
\cite{p2}). Requiring this zero mode to vanish as a further gauge condition,
one finds $e_0{}^+=1$ and $e_0{}^-=\o_0=0$.  In other gauges the
Lagrange multipliers can become also non trivial quantum
operators. Furthermore, it is a special feature of the chosen
gauge that the obtained operators are $x^1$-independent.  (The
existence of this gauge shows that $B_3 = const$ is an isometry
or Killing direction of the metric). Again different choices of
gauge conditions are interpreted as corresponding to different
types of questions or measuring devices.

The alternative, at least for the paramtrization invariant NRP
equivalent  procedure to reintroduce time
within the quantum theory
was the direct implementation of the gauge in the wave
functions. For this it was decisive that the initially chosen
polarization of the wave functions contained the phase space
variable subject to the gauge.  To implement \re{eich})
analogously within the gravity theory under consideration, we
Fourier transform \re{psi}), multiplied by $\d[\6 Q_{(2)}]$,
with respect to $B_-(x^1)$. The result is
\begin{eqnarray}
\lefteqn{ \exp \left({i \0 \hbar} \oint [E_3 \6 \varphi +
{\6 B_+ B_3+ [{\b^2 \0 3} (B_3)^3 -\g^2 tr(E^2)B_3]e_1{}^- \0
B_+}] dx^1 \right)}
\hspace{6cm} &&
\nn
&& \Pi_{x^1}\left({const \0  B_+}\right) \hat
\Psi_n(p_{(2)}), \label{psiha} \end{eqnarray}
in which $\hat \Psi_n$ is the Fourier  transform of the ordinary
function $\ti \Psi_n$.  Eq. \re{psiha})  certainly is in
agreement with the general solution of the quantum constraints
in a $(B_+,B_3,e_1{}^-,E)$ representation. In the gauge
\re{eich})  the quantum wave functions take the form
\be  \sum_n \exp{\left[{-i \0 \hbar} \left({\g^2n^2 \0 8}\t +{\b^2\pi^2 \0
3 p_{(2)}^2}\t^3 \right)\right]}  c_n(p_{(2)}) |n\rangle,
\la{time}  \ee where $|n\rangle$ denotes the eigenfunctions of
$q_{(1)}$ (inclusive $\exp[(i /\hbar) \oint E_3 \6 \varphi
dx^1]$) and we have reabsorbed the divergent factor of
\re{psiha}), being a function of $p_{(2)}$, into $c_n(p_{(2)})$.

At this point the case $\b=0$ is of special interest: for it
$S_H$ is seen to describe a Yang Mills theory coupled to a {\em
flat}  metric. Thus in some sense it is the parametrization
(i.e.\ diffeomorphism) invariant formulation of the usual   Yang
Mills theory on the cylinder (with rigid Minkowski background
metric). If we ignore the $p_{(2)}$ dependence of $c_n$ for a
moment, \re{time}) with $\b=0$ indeed coincides with the time
evolution generated by the (nonvanishing) Yang Mills Hamiltonian
$-\g^2 \oint tr E^2 dx^1 \equiv  \g^2 \pi q_{(1)}$.  This
agreement gives support to the method used to derive \re{time}).

The reason for the $p_{(2)}$-dependence of $c_n$ is due to the
fact that in the formulation \re{SH}) with $\b=0$ the metric
induced circumference of the cylinder became a dynamical
variable (on shell one has $p_{(2)} \propto
\oint_{B_3=const} \sqrt{g_{11}} dx^1$).
Within \re{eich}) one finds $-\oint G_+ \sim H$ to effectively
implement the Schroedinger equation corresponding to \re{time}).
The effective Hamiltonian acting on $c_n  |n\rangle$ is
$-(\g^2/2) \oint tr E^2 dx^1 - \b^2\pi^2 \t^2/p_{(2)}^2$.  Thus
generically the above procedure yields time dependent
Hamiltonians (cf.\ also \cite{p2}).

The strategies developed at the example of a NRP to resolve the
'issue of time' within a quantum theory of gravity produced
sensible results for the toy model \re{S}). They, however,
relied heavily on either the knowledge of all Dirac observables
or on some specifically chosen polarization. To cope with the
considerable technical difficulties of a quantum theory of four
dimensional gravity, it might be worthwhile to extend the
applicability of the method. One way to do so within our model
is to allow for equivalence classes of wave functions coinciding
at $\6 Q_{(2)}=0$, the latter condition being enforced  within
the inner product \cite{p2}.  In this way one can, e.g.,
implement the gauge condition $\6 e_1{}^-=0$ as an operator
condition in the $B$ polarization of the wave functions as well,
whereas a straightforward implementation of $\oint
e_1{}^-=const$  seems again inadmissible.

\end{document}